\documentclass[12pt,a4paper]{article}

\usepackage{pslatex}	% fit more on page using PS-fonts 
\usepackage{amsfonts}
\usepackage{amsmath}
\usepackage{amsthm}
\usepackage{cite}
\usepackage{epsf}

\def\beq{\begin{equation}}
\def\eeq{\end{equation}}
\def\bea{\begin{eqnarray}}
\def\eea{\end{eqnarray}}
\let\nn=\nonumber
\def\beann{\begin{eqnarray*}}
\def\eeann{\end{eqnarray*}}

\newcommand{\append}[1]{\protect\stepcounter{section}
			\setcounter{equation}{0} 
			\section*{Appendix \thesection : #1}
			\addcontentsline{toc}{section}{Appendix
			\thesection: #1}}

\let\a=\alpha \let\be=\beta \let\g=\gamma \let\de=\delta

  \let\la=\lambda \let\m=\mu
  \let\p=\pi  
 
  \let\PH=\Phi 
\let\Om=\Omega  
\let\La=\Lambda  

\let\qd=\quad \let\qqd=\qquad 

\def\tst#1{{\textstyle #1}}

\theoremstyle{plain}

\newtheorem{lemma}{Lemma}
\newtheorem{corollary}{Corollary}
\newtheorem*{corollary*}{Corollary}

\theoremstyle{definition}
\newtheorem*{remark}{Remark}

\def\2{\frac{1}{2}} \def\4{\frac{1}{4}}

\def\6{\partial}

\def\<{\langle} \def\>{\rangle}

\let\auf=\uparrow \let\ab=\downarrow

\def\CH{{\cal H}} \def\CL{{\cal L}}

\def\CT{{\cal T}}

\def\i{{\rm i}}

 \def\End{{\rm End}} \def\id{{\rm id}}

\def\res{{\rm res}}

\def\tr{{\rm tr}} 
\def\str{{\rm str}}

\begin{document}

\thispagestyle{empty}

\begin{center}

{\Large {\bf Algebraic Bethe ansatz for the gl(1$|$2) generalized
model and Lieb-Wu equations \\}}

\vspace{7mm}

{\large Frank G\"{o}hmann\footnote[2]{%
e-mail: Frank.Goehmann@uni-bayreuth.de}\\

\vspace{5mm}

Theoretische Physik I, Universit\"at Bayreuth, 95440 Bayreuth,
Germany\\}

\vspace{20mm}

{\large {\bf Abstract}}

\end{center}

\begin{list}{}{\addtolength{\rightmargin}{10mm}
               \addtolength{\topsep}{-5mm}}
\item
We solve the gl(1$|$2) generalized model by means of the algebraic
Bethe ansatz. The resulting eigenvalue of the transfer matrix and the
Bethe ansatz equations depend on three complex functions, called the
parameters of the generalized model. Specifying the parameters
appropriately, we obtain the Bethe ansatz equations of the
supersymmetric $t$-$J$ model, the Hubbard model, or of Yang's model of
electrons with delta interaction. This means that the Bethe ansatz
equations of these (and many other) models can be obtained from a
common algebraic source, namely from the Yang-Baxter algebra generated
by the gl(1$|$2) invariant $R$-matrix.
\\[2ex]
{\it PACS: 05.50.+q, 71.10.Fd, 71.10.Pm}
\end{list}

\clearpage

\section{Introduction}
An important class of exactly solvable 1+1 dimensional quantum
systems is related to the Yang-Baxter algebra. The theory of these
systems is called the quantum inverse scattering method \cite{KBIBo}.
The central object of the quantum inverse scattering method is the
monodromy matrix, whose elements generate the Yang-Baxter algebra, and
whose trace, the transfer matrix, determines a lattice statistical
model, which, in many cases, is of interest in physics. The structure
of the Yang-Baxter algebra is fixed by a set of numerical functions,
the elements of the $R$-matrix, which satisfies the famous Yang-Baxter
equation \cite{Jimbo89}.

Solving an exactly solvable model means at first instance to calculate
the spectrum and the eigenfunctions of the transfer matrix. In many
cases of interest this task can be accomplished using the quadratic
commutation relations between the elements of the monodromy matrix.
The model is then said to be solved by algebraic Bethe ansatz. For the
algebraic Bethe ansatz to work it has to be possible to identify the
elements of the monodromy matrix as `particle' creation and annihilation
operators. In particular, a pseudo vacuum state must exist, which
is annihilated by all the annihilation operators. In all known cases,
where an algebraic Bethe ansatz was successful so far, the elements of
the monodromy matrix could be arranged in such a way, that the monodromy
matrix acts as an upper triangular matrix on the pseudo vacuum, and
the pseudo vacuum is an eigenstate of its diagonal elements. Still,
even if a pseudo vacuum exists, on which the monodromy matrix acts
triangularly, no general recipe for an algebraic Bethe ansatz is known,
and the remaining calculations may be rather involved (see \cite{MaRa97}
for a highly non-trivial example).

The algebraic Bethe ansatz is best understood for models with the
$R$-matrix of the spin-$\2$ XXX and XXZ chains \cite{KBIBo}. In this
case there is only one monodromy matrix element below the diagonal,
which must annihilate the pseudo vacuum. The particular form of the
vacuum eigenvalues of the diagonal elements of the monodromy matrix is
not used in the construction of the transfer matrix eigenvectors. The
vacuum eigenvalues, usually denoted $a(\la)$ and $d(\la)$, enter the
algebraic Bethe ansatz solution as functional parameters. Thus, one may
think of a model {\em defined} by the functional parameters and the
triangular action of the monodromy matrix on the pseudo vacuum. The
question whether such kind of `generalized model' can be realized for
arbitrary parameters $a(\la)$ and $d(\la)$ was first addressed in
\cite{Korepin82b} and after some refinement was answered
affirmatively in \cite{Tarasov84,Tarasov85}. For the models with XXX
and XXZ $R$-matrix the arbitrariness of the functional parameters was
the key tool in Korepin's proof \cite{Korepin82} of the norm formula and
in Slavnov's work \cite{Slavnov89} on form factors.

The simplest models which allow for a {\em nested} algebraic Bethe
ansatz are the models with gl(n) invariant $R$-matrix \cite{KuSk80,%
KuRe82}. Considering the fundamental representations of these models
one observes that not only the monodromy matrix elements below the
diagonal annihilate the pseudo vacuum, but additional zeros appear
above the diagonal \cite{KuRe81}. This fact simplifies the algebraic
Bethe ansatz for the fundamental representation as compared to the more
general case, where the action of all the elements of the monodromy matrix
above the diagonal is non-trivial. For the solution of this more
general case a new concept, the vacuum subspace, was introduced by
Kulish and Reshetikhin \cite{KuRe83}. This new concept enabled to
perform the algebraic Bethe ansatz for the models with gl(n) invariant
$R$-matrix with the same generality as in the gl(2) case. The resulting
eigenvalue of the transfer matrix and the Bethe ansatz equations depend
on $n$ functional parameters $a_1 (\la), \dots, a_n (\la)$, which,
together with the triangular action of the monodromy matrix on the
pseudo vacuum, define the so-called gl(n) generalized model
\cite{Reshetikhin86}. Considering the parameters $a_1 (\la), a_2 (\la),
a_3 (\la)$ as free parameters Reshetikhin derived the norm formula for
the gl(3) generalized model \cite{Reshetikhin86}.

In this work we construct the algebraic Bethe ansatz solution for
the gl(1$|$2) generalized model. This work has three motivations:
\begin{enumerate}
\item
To perform the algebraic Bethe ansatz for all models with gl(1$|$2)
invariant graded $R$-matrix \cite{KuSk80,Kulish85}.
\item
To pave the ground for the calculation of the norm for this class of
models.
\item
To point out an interesting relation to the Hubbard model, namely, that
the Lieb-Wu equations \cite{LiWu68} can be generated by the Yang-Baxter
algebra with gl(1$|$2) invariant $R$-matrix.
\end{enumerate}
Let us comment here on point (i) above. Points (ii) and (iii) will be
further discussed in section 7.

Numerous articles have appeared about models connected to the
Yang-Baxter algebra generated by the gl(1$|$2) invariant $R$-matrix.
In first place we have to mention the lattice gas model of Lai and
Sutherland \cite{Lai74,Sutherland75} and the supersymmetric $t$-$J$
model \cite{Schlottmann87}. Their relation to the graded Yang-Baxter
algebra was explored in \cite{EsKo92,FoKa93}. Both models are
connected to the fundamental representation of gl(1$|$2). Their
algebraic Bethe ansatz solution was obtained in \cite{Kulish85} (see
also \cite{EsKo92}). More recently there was a wave of interest in
models related to higher dimensional representations of gl(1$|$2)
\cite{SNR77,Marcu80}. A fermionic model related to the four-dimensional
typical representation was introduced in \cite{Karnaukhov94,%
Maassarani95,BGLZ95} and was named supersymmetric U model. Algebraic
Bethe ansatz solutions of this model were obtained in \cite{PfFr96,%
PfFr97,RaMa96} (see also \cite{HGL96}). An impurity $t$-$J$ model,
where the fundamental representation at one lattice site is replaced
by the typical four-dimensional representation, was considered in
\cite{BEF97,FLT99}. Finally, the latest example in our list is the family
of `doped Heisenberg chains' \cite{Frahm99} related to the atypical
representations of gl(1$|$2).

The algebraic Bethe ansatz solutions of all the above mentioned models
can be obtained as special cases of the solution presented in this
work. In this sense, we claim that we performed the algebraic Bethe
ansatz for all models with gl(1$|2$) invariant $R$-matrix. We would
also like to point out that our result applies to the Bethe ansatz for
the quantum transfer matrix of the supersymmetric $t$-$J$ model and can
be used to prove the conjectures, presented in \cite{JKS97,KWZ97}, about
the quantum transfer matrix eigenvalue and the corresponding Bethe
ansatz equations.

Let us note that one of the Bethe ansatz solutions in our above list is
quite different from the others and is not covered by our approach.
Ramos and Martins' solution \cite{RaMa96} is based on the Yang-Baxter
algebra generated by the intertwiner of two typical four-dimensional
representations of gl(1$|$2) and not on the fundamental $R$-matrix.
They obtain the same result for the eigenvalue of the transfer matrix
of the supersymmetric U model and the same Bethe ansatz equations as
in \cite{PfFr96}, but their construction of the eigenvectors is quite
different. It will be interesting to find out the relation between
the two seemingly different sets of eigenvectors.

The plan of this work is the following: In section 2 we make more
precise what we mean by the gl(1$|$2) generalized model. In preparation
of the algebraic Bethe ansatz we work out the block structure of
the Yang-Baxter algebra in section 3. Section 4 is the core of this
work. It is devoted to the algebraic Bethe ansatz. In section 5
we consider the fundamental graded representation, related to the
supersymmetric $t$-$J$ model. This section has been included to give
an example of how our slightly abstract formalism relates back to
physics. Following the example the reader will be able to work out
by himself the cases he is interested in. In section 6 we point out
a relation of our general result to the Lieb-Wu equations. Section 7
is devoted to a discussion of possible applications.

In order not to overload this article with formulae we shall restrict
ourselves to a fixed choice $(+--)$ of the grading. The other two
possible cases (see e.g.\ \cite{Kulish85,EsKo92}) are left to the
reader. They present no more difficulty than the case considered in
the text.

\section{The gl(1$|$2) generalized model}
The starting point of our considerations is the gl(1$|$2) invariant
rational $R$-matrix \cite{KuSk80,Kulish85} with matrix elements
\begin{equation} \label{rm}
     R^{\a \g}_{\be \de} (\la) = a(\la) (-1)^{p(\a) p(\g)}
                                 \de^\a_\be \de^\g_\de +
				 b(\la) \de^\a_\de \de^\g_\be \qd,
\end{equation}
$\a, \be, \g, \de = 1, 2, 3$. For simplicity we restrict ourselves to
the grading $p(1) = 0$, $p(2) = p(3) = 1$. The matrix $R(\la)$ solves
the Yang-Baxter equation and is invariant with respect to the action
of the fundamental representation of gl(1$|$2). It satisfies the
compatibility condition \cite{KuSk80}
\begin{equation} \label{comp}
     R^{\a \g}_{\be \de} (\la) = (-1)^{p(\a) + p(\be) + p(\g) + p(\de)}
                                 R^{\a \g}_{\be \de} (\la) \qd.
\end{equation}
The complex valued functions $a(\la)$ and $b(\la)$ of the spectral
parameter $\la \in {\mathbb C}$ are defined as
\begin{equation} \label{defab}
     a(\la) = \frac{\la}{\la + \i c} \qd, \qd
     b(\la) = \frac{\i c}{\la + \i c} \qd.
\end{equation}
They satisfy the equation $a(\la) + b(\la) = 1$. The additional complex
parameter $c$ will be called the coupling constant. We shall further
need the matrix $\check R (\la)$ defined by
\begin{equation}
     \check R^{\a \g}_{\be \de} (\la) = R^{\g \a}_{\be \de} (\la) \qd.
\end{equation}

The graded Yang-Baxter algebra with $R$-matrix $R(\la)$ is the graded,
associative algebra (with unit element) generated by the elements
$\CT^\a_\be (\la)$, $\a, \be = 1, 2, 3$, of the so-called monodromy
matrix modulo the relations
\begin{equation} \label{gyba}
     \check R (\la - \m) \big( \CT (\la) \otimes_s \CT (\m) \big) =
     \big( \CT (\m) \otimes_s \CT (\la) \big) \check R (\la - \m) \qd.
\end{equation}
We assume that the elements of the monodromy matrix are of definite
parity, $\p(\CT^\a_\be (\la)) = p(\a) + p(\be)$. The symbol $\otimes_s$
denotes the super tensor product associated with the grading $p(1) = 0$,
$p(2) = p(3) = 1$. For a definition see appendix~A.

The gl(1$|$2) generalized model is the set of all (linear)
representations of the graded Yang-Baxter algebra (\ref{gyba}) with
highest vector $\Om$, defined by the fact that $\CT (\la)$ acts
triangularly on $\Om$,
\begin{align} \label{para}
     & \CT^1_1 (\la) \Om = a_1 (\la) \Om \qd, \qd
       \CT^2_2 (\la) \Om = a_2 (\la) \Om \qd, \qd
       \CT^3_3 (\la) \Om = a_3 (\la) \Om \qd, \\[1ex] \label{null}
     & \CT^\a_\be (\la) \Om = 0 \qd, \qd \text{for $\a > \be$} \qd.
\end{align}
The complex valued functions $a_j (\la)$, $j = 1, 2, 3$, are called the
parameters of the generalized model. These parameters characterize
the representation in a similar manner as the highest weight in a
highest weight representation of a Lie algebra.

Let us denote the representation space of a given representation of the
generalized model by $\CH$. It is clear from the quadratic commutation
relations contained in the graded Yang-Baxter algebra (\ref{gyba}) and
from (\ref{para}), (\ref{null}) that we may assume that $\CH$ is spanned
by all vectors of the form
\begin{equation}
     \PH (\la_1, \dots, \la_N) = \CT^{\a_1}_{\be_1} (\la_1) \dots
                                 \CT^{\a_N}_{\be_N} (\la_N) \Om \qd,
\end{equation}
where $\a_k < \be_k$, $k = 1, \dots, N$. This assumption is at least
sensible for a finite dimensional representation space $\CH$.

Let us define the transfer matrix
\begin{equation}
     t(\la) = (-1)^{p(\a)} \CT^\a_\a (\la) = \str (\CT (\la)) \qd.
\end{equation}
Since $\check R (\la)$ is invertible for generic values of $\la \in
{\mathbb C}$, we conclude from (\ref{comp}) and (\ref{gyba}) that
\begin{equation}
     [t(\la), t(\m)] = 0
\end{equation}
for all generic $\la, \m \in {\mathbb C}$. Thus $t(\la)$ and $t(\m)$
have a common system of eigenfunctions. In other words, the
eigenvectors of $t(\la)$ are independent of the spectral parameter $\la$.
The task of the algebraic Bethe ansatz for the generalized model is
to diagonalize $t(\la$), i.e., to solve the eigenvalue problem
\begin{equation}
     t(\la) \PH = \La (\la) \PH \qd.
\end{equation}
It is a remarkable fact that this task can be accomplished by solely
using the graded Yang-Baxter algebra (\ref{gyba}) and the properties
(\ref{para}) and (\ref{null}) of the highest vector $\Om$. In particular,
it is {\em not} necessary to require that $\CT^2_3 (\la) \Om = 0$ as
in case of the fundamental graded representation, which corresponds to
the supersymmetric $t$-$J$ model.

\section{The graded Yang-Baxter algebra}
The structure of the graded Yang-Baxter algebra (\ref{gyba}) becomes
much clearer after rewriting it in block form. We introduce the
shorthand notations
\begin{equation}
\begin{split}
     & B(\la) = \big( B_1 (\la), B_2 (\la) \big) \qd, \qd
       C(\la) = \begin{pmatrix} C^1 (\la) \\ C^2 (\la) \end{pmatrix} \qd,
                \\
     & D(\la) = \begin{pmatrix} D^1_1 (\la) & D^1_2 (\la) \\
                                D^2_1 (\la) & D^2_2 (\la) \end{pmatrix}
				\qd.
\end{split}
\end{equation}
Then we can write the $3 \times 3$ monodromy matrix $\CT (\la)$ as
\begin{equation}
     \CT (\la) = \begin{pmatrix} A (\la) & B_1 (\la) & B_2 (\la) \\
                                 C^1 (\la) & D^1_1 (\la) & D^1_2 (\la) \\
				 C^2 (\la) & D^2_1 (\la) & D^2_2 (\la)
				 \end{pmatrix} =
                 \begin{pmatrix} A (\la) & B (\la) \\
		                 C (\la) & D (\la)
				 \end{pmatrix} \qd.
\end{equation}
The defining relations of the graded Yang-Baxter algebra (\ref{gyba})
can be thought of as a $9 \times 9$ matrix equation. Let us denote the
$n \times n$ unit matrix by $I_n$. A similarity transformation with the
matrix
\begin{equation}
     X = \begin{pmatrix} I_4 & & \\
            & \begin{pmatrix} 0 & 0 & 1 \\ 1 & 0 & 0 \\ 0 & 1 & 0
	      \end{pmatrix} & \\
	    & & I_2
         \end{pmatrix} \qd,
\end{equation}
which cyclically permutes the 5th, 6th and 7th row, transforms this
$9 \times 9$ equation into
\begin{multline} \label{blockyba}
     \begin{pmatrix}
        1 & & & \\
	& b \, I_2 & a \, I_2 & \\
	& a \, I_2 & b \, I_2 & \\
	& & & \check r
     \end{pmatrix}
     \begin{pmatrix}
        A \otimes \bar A & A \otimes \bar B &
        B \otimes \bar A & B \otimes \bar B \\
        A \otimes \bar C & A \otimes \bar D &
        - B \otimes \bar C & - B \otimes \bar D \\
        C \otimes \bar A & C \otimes \bar B &
        D \otimes \bar A & D \otimes \bar B \\
        - C \otimes \bar C & - C \otimes \bar D &
        D \otimes \bar C & D \otimes \bar D
     \end{pmatrix} = \\[1ex]
     \begin{pmatrix}
        \bar A \otimes A & \bar A \otimes B &
        \bar B \otimes A & \bar B \otimes B \\
        \bar A \otimes C & \bar A \otimes D &
        - \bar B \otimes C & - \bar B \otimes D \\
        \bar C \otimes A & \bar C \otimes B &
        \bar D \otimes A & \bar D \otimes B \\
        - \bar C \otimes C & - \bar C \otimes D &
        \bar D \otimes C & \bar D \otimes D
     \end{pmatrix}
     \begin{pmatrix}
        1 & & & \\
	& b \, I_2 & a \, I_2 & \\
	& a \, I_2 & b \, I_2 & \\
	& & & \check r
     \end{pmatrix} \qd.
\end{multline}
For the formula to fit on the line we suppressed the arguments and
adopted the following convention: $A, \dots, D$ depend on $\la$, and
a bar means here that $\la$ is replaced by $\m$. Furthermore, $a =
a(\la - \m)$ and $b = b(\la - \m)$. The $4 \times 4$ matrix
\begin{equation}
     \check r = \begin{pmatrix} b - a & & & \\ & b & - a & \\
                                & - a & b & \\ & & & b - a
                \end{pmatrix}
\end{equation}
is of `six-vertex form'. It satisfies the Yang-Baxter equation and is
unitary,
\begin{equation}
     \check r (\la) \check r (- \la) = I_4 \qd.
\end{equation}
We would like to remark that the defining relations of the graded
Yang-Baxter algebra of the gl(1$|$2) model, when written in block form
(\ref{blockyba}), resemble the corresponding relations for the gl(1$|$1)
model.

Out of the 16 relations contained in (\ref{blockyba}) we shall need the
following 4 for the algebraic Bethe ansatz,
\begin{align} \label{zamo}
     & B(\la) \otimes B(\m) = \big( B(\m) \otimes B(\la) \big)
                              \check r (\la - \m) \qd, \\[1ex] \label{ab}
     & A(\la) \otimes B(\m) = - \, \frac{b(\m - \la)}{a(\m - \la)} \,
                              B(\la) \otimes A(\m) +
			      \frac{B(\m) \otimes A(\la)}{a(\m - \la)}
			      \qd, \\[1ex] \label{db}
     & D(\la) \otimes B(\m) = \frac{b(\la - \m)}{a(\la - \m)} \,
                              B(\la) \otimes D(\m) -
			      \big( B(\m) \otimes D(\la) \big)
			      \, \frac{\check r (\la - \m)}{a(\la - \m)}
			      \qd, \\[1ex] \label{dyba}
     & \check r (\la - \m) \big( D(\la) \otimes D(\m) \big) =
       \big( D(\m) \otimes D(\la) \big) \check r (\la - \m) \qd.
\end{align}
Note that, by (\ref{zamo}), $B(\la)$ constitutes a representation of the
Zamolodchikov algebra, and, by (\ref{dyba}), $D(\la)$ is a representation
of the Yang-Baxter algebra of the gl(2) model.

\section{The algebraic Bethe ansatz}
Our goal is to calculate the eigenvectors of the transfer matrix $t(\la) =
A(\la) - \tr(D(\la))$. In analogy with the gl(1$|$1) case we shall
first of all consider the commutation relations of a multiple
tensor product $B(\la_1) \otimes \dots \otimes B(\la_N)$ with $A(\la)$
and $\tr(D(\la))$. These commutation relations can be obtained be
iterating equations (\ref{ab}) and (\ref{db}). We first present the
result of the iteration and explain our notation below,
\begin{align} \label{abmult}
     & A(\la) \Bigl[ \bigotimes_{j=1}^N B(\la_j) \Bigr] = 
          \Bigl[ \bigotimes_{j=1}^N B(\la_j) \Bigr]
	  A(\la) \prod_{j=1}^N \frac{1}{a(\la_j - \la)} \nn \\
        & \qqd - \sum_{j=1}^N \biggl\{ B(\la) \otimes
	        \Bigl[ \bigotimes_{k=1 \atop k \ne j}^N B(\la_k) \Bigr]
		\biggr\} S_{j-1} \,
		A(\la_j) \, \frac{b(\la_j - \la)}{a(\la_j - \la)}
		\prod_{k=1 \atop k \ne j}^N \frac{1}{a(\la_k - \la_j)}
		\qd, \\[1ex] \label{dbmult}
     & D(\la) \otimes \Bigl[ \bigotimes_{j=1}^N B(\la_j) \Bigr] = 
        (-1)^N \biggl\{
	   I_2 \otimes \Bigl[ \bigotimes_{j=1}^N B(\la_j) \Bigr] \biggr\}
	   \tilde T (\la) \prod_{j=1}^N \frac{1}{a(\la - \la_j)} \nn \\
        & \qqd - (-1)^N \sum_{j=1}^N \biggl\{ I_2 \otimes B(\la) \otimes
	        \Bigl[ \bigotimes_{k=1 \atop k \ne j}^N B(\la_k) \Bigr]
		\biggr\} P_{01} S_{j-1}^{(0)} \cdot \nn \\
        & \mspace{242.0mu} \cdot \tr \bigl(\tilde T (\la_j)\bigr)
		\, \frac{b(\la - \la_j)}{a(\la - \la_j)}
		\prod_{k=1 \atop k \ne j}^N \frac{1}{a(\la_j - \la_k)}
		\qd.
\end{align}
Here the operators $B(\la)$ in the multiple tensor products are multiplied
in ascending order. We make use of the usual conventions for the embedding
of linear operators into tensor product spaces and define
\begin{equation}
     \check r_{j-1, j} (\la) = I_2^{\otimes (j-2)} \otimes \check r (\la)
                               \otimes I_2^{\otimes (N-j)}
\end{equation}
for $j = 2, \dots, N$ as an operator in $\bigl( \End({\mathbb
C}^2) \bigr)^{\otimes N}$. The operators $S_{j-1}$ appearing on the
right hand side of (\ref{abmult}) are then defined as
\begin{equation}
     S_{j-1} = \check r_{1, 2} (\la_1 - \la_j)
               \check r_{2, 3} (\la_2 - \la_j) \dots
               \check r_{j-1, j} (\la_{j-1} - \la_j) \qd,
\end{equation}
for $j = 2, \dots, N$. We further define $S_0 = \id$. Then the notation
used in (\ref{abmult}) is explained.

It remains to explain the operators $P_{01}$, $S_{j-1}^{(0)}$ and
$\tilde T (\la)$ appearing at the right hand side of equation
(\ref{dbmult}). These operators act in one more auxiliary space `zero'.
\begin{equation}
     S_{j-1}^{(0)} = I_2 \otimes S_{j-1} \qd, \qd
     P_{01} = P \otimes I_2^{\otimes (N-1)} \qd,
\end{equation}
where $P$ is the permutation matrix on ${\mathbb C}^2 \otimes
{\mathbb C}^2$ with matrix elements $P^{\a \g}_{\be \de} = \de^\a_\de
\de^\g_\be$. In order to define $\tilde T (\la)$ we introduce the
gl(2) $R$-matrix
\begin{equation}
     r(\la) = P \check r (\la)
\end{equation}
and the monodromy matrix of the corresponding fundamental inhomogeneous
model,
\begin{equation} \label{deftaux}
     T^{(0)} (\la) = r_{0, N}^{(0)} (\la - \la_N) \dots
                     r_{0, 1}^{(0)} (\la - \la_1) \qd.
\end{equation}
We shall interpret this auxiliary monodromy matrix as a $2 \times 2$
matrix in space zero with entries acting on spaces $1, \dots, N$. We
further introduce
\begin{equation}
     D_0^{(0)} (\la) = D(\la) \otimes I_2^{\otimes N} \qd.
\end{equation}
Eventually, $\tilde T(\la)$ is defined as
\begin{equation} \label{deftt}
     \tilde T (\la) = D_0^{(0)} (\la) T^{(0)} (\la) \qd.
\end{equation}
This expression, too, can be understood as a $2 \times 2$ matrix, the
trace of which appears under the sum on the right hand side of
equation (\ref{dbmult}).

Equations (\ref{abmult}) and (\ref{dbmult}) were obtained by the
usual symmetry arguments of the algebraic Bethe ansatz. Their derivation
solely relies on (\ref{zamo})-(\ref{db}). As an immediate consequence
of (\ref{zamo}) we obtain the equation
\begin{equation} \label{bbmult}
     \bigotimes_{k=1}^N B(\la_k) = \biggl\{ B(\la_j) \otimes \Bigl[
                                   \bigotimes_{k=1 \atop k \ne j}^N
				   B(\la_k) \Bigr] \biggr\} S_{j-1} \qd,
\end{equation}
which was repeatedly used in the derivation of (\ref{abmult}),
(\ref{dbmult}). A mathematically rigorous proof of (\ref{abmult}),
(\ref{dbmult}), which is not too hard to do, may be obtained by
induction over $N$. Some useful identities needed in the proof of the
less trivial case of (\ref{dbmult}) are provided in appendix B.

A commutation relation, appropriate for our purposes, of the transfer
matrix $t(\la)$ with a multiple tensor product $B(\la_1) \otimes \dots
\otimes B(\la_N)$ now easily follows from (\ref{abmult}) and
(\ref{dbmult}). We have to take the trace of equation (\ref{dbmult}) in
space zero and have to subtract the result from equation (\ref{abmult}).
Using the fact that $b(\la)/a(\la) = \i c/\la$ is an odd function of
$\la$ we obtain
\begin{align} \label{tbmult}
     & t(\la) \Bigl[ \bigotimes_{j=1}^N B(\la_j) \Bigr] =
        \Bigl[ \bigotimes_{j=1}^N B(\la_j) \Bigr]
	\nn \\ & \mspace{168.0mu} \cdot
	  \Bigl\{
	     A(\la) \prod_{j=1}^N \frac{1}{a(\la_j - \la)} - (-1)^N
	     \tr \bigr( \tilde T (\la) \bigl)
	     \prod_{j=1}^N \frac{1}{a(\la - \la_j)}
	  \Bigr\} \nn \\[1ex]
        & \qd + \sum_{j=1}^N \biggl\{ B(\la) \otimes
	        \Bigl[ \bigotimes_{k=1 \atop k \ne j}^N B(\la_k) \Bigr]
		\biggr\} S_{j-1} \, \frac{b(\la - \la_j)}{a(\la - \la_j)}
		\prod_{k=1 \atop k \ne j}^N \frac{1}{a(\la_j - \la_k)}
		\nn \\
        & \mspace{168.0mu} \cdot
		\Bigl\{
		   A(\la_j) \prod_{k=1 \atop k \ne j}^N
		   \frac{a(\la_j - \la_k)}{a(\la_k - \la_j)} +
		   (-1)^N \tr \bigl(\tilde T (\la_j)\bigr)
		\Bigr\} \qd.
\end{align}

Before proceeding with the Bethe ansatz calculation we have to gain
more insight into the mathematical structure of the various expressions
in equation (\ref{tbmult}): The vector $B(\la_1) \otimes \dots \otimes
B(\la_N)$ may be thought of as a $2^N$ component row vector with entries
acting on the representation space $\CH$. The components of this vector
are
\begin{equation}
     \Bigl[ \bigotimes_{j=1}^N B(\la_j) \Bigr]_{i_1, \dots, i_N} =
     B_{i_1} (\la_1) \dots B_{i_N} (\la_N) \qd,
\end{equation}
$i_n = 1, 2$. Equivalently, $B(\la_1) \otimes \dots \otimes B(\la_N)$
is a vector in $({\mathbb C}^2)^{\otimes N} \otimes \End(\CH)$. Suppose
$\hat F^{i_1, \dots, i_N} \in \CH$ for all $i_1, \dots, i_N = 1, 2$.
Then $\hat F$ can be defined as a column vector with $2^N$ components
$\hat F^{i_1, \dots i_N}$ that are vectors in $\CH$, or, equivalently,
$\hat F \in ({\mathbb C}^2)^{\otimes N} \otimes \CH$. It follows that
\begin{equation}
     \Bigl[ \bigotimes_{j=1}^N B(\la_j) \Bigr] \hat F =
        B_{i_1} (\la_1) \dots B_{i_N} (\la_N) \hat F^{i_1, \dots, i_N}
        \in \CH \qd.
\end{equation}
Following Kulish and Reshetikhin \cite{KuRe83} let us define the
`vacuum subspace' $\CH_0 \subset \CH$ by the conditions
\begin{align} \label{h01}
     & A(\la) \PH = a_1 (\la) \PH \qd, \\
     & C(\la) \PH = 0 \qd, \label{h02}
\end{align}
for all $\PH \in \CH_0$. Clearly, $\CH_0$ is a linear subspace of $\CH$.
The following lemma holds~\cite{Reshetikhin86}.
\begin{lemma}
$\CH_0$ is invariant under the action of $D(\la)$.
\end{lemma}
\begin{proof}
From the Yang-Baxter algebra written in block form (\ref{blockyba}) we
deduce the equations
\begin{align} \label{bcda}
     - & b(\la - \m) \, B(\la) \otimes C(\m)
        + a(\la - \m) \, D(\la) \otimes A(\m) = \nn \\ & \mspace{124.0mu}
          a(\la - \m) \, A(\m) \otimes D(\la)
        - b(\la - \m) \, B(\m) \otimes C(\la) \qd, \\[1ex] \label{rdc}
       & \check r (\la - \m) \bigl( D(\la) \otimes C(\m) \bigr) =
        \nn \\ & \mspace{106.0mu}
        - a(\la - \m) \, C(\m) \otimes D(\la)
	+ b(\la - \m) \, D(\m) \otimes C(\la) \qd.
\end{align}
Let $\PH \in \CH_0$. Acting with (\ref{bcda}) and (\ref{rdc}) on $\PH$
we obtain
\begin{align}
     \bigl( A(\m) \otimes D(\la) \bigr) \PH & = a_1 (\m) D(\la) \PH
                                                \qd, \\
     \bigl( C(\m) \otimes D(\la) \bigr) \PH & = 0 \qd,
\end{align}
and the lemma is proven.
\end{proof}

\begin{corollary}
The space spanned by all linear combinations of vectors of the form
$D^1_2 (\m_1) \dots D^1_2 (\m_M) \, \Om$ is a linear subspace of $\CH_0$.
\end{corollary}

Thanks to our lemma we can now proceed with the so-called second level
Bethe ansatz. We introduce the notation
\begin{equation}
     T^{(0)} (\la) = \begin{pmatrix} A^{(0)} (\la) & B^{(0)} (\la) \\
                                     C^{(0)} (\la) & D^{(0)} (\la)
                     \end{pmatrix} \qd, \qd
     \tilde T (\la) = \begin{pmatrix} \tilde A (\la) & \tilde B (\la) \\
                                      \tilde C (\la) & \tilde D (\la)
                      \end{pmatrix} \qd.
\end{equation}
Then, by the definition (\ref{deftt}) of $\tilde T (\la)$,
\begin{equation} \label{abcdt}
\begin{split}
     \tilde A (\la) = A^{(0)} (\la) D^1_1 (\la) +
                      C^{(0)} (\la) D^1_2 (\la) \qd, \\
     \tilde B (\la) = B^{(0)} (\la) D^1_1 (\la) +
                      D^{(0)} (\la) D^1_2 (\la) \qd, \\
     \tilde C (\la) = A^{(0)} (\la) D^2_1 (\la) +
                      C^{(0)} (\la) D^2_2 (\la) \qd, \\
     \tilde D (\la) = B^{(0)} (\la) D^2_1 (\la) +
                      D^{(0)} (\la) D^2_2 (\la) \qd.
\end{split}
\end{equation}
Here we used the fact that
\begin{equation} \label{dtcom}
     [D^\a_\be (\la), {T^{(0)}}^\g_\de (\la)] = 0 \qd, \qd
       \text{for $\a, \be, \g, \de = 1, 2$.}
\end{equation}
Since both, $D(\la)$ and $T^{(0)} (\la)$, are representations of the
Yang-Baxter algebra with $R$-matrix $r (\la)$ (see (\ref{dyba})),
equation (\ref{dtcom}) allows us to conclude that
\begin{equation} \label{gl2yba}
     \check r (\la - \m)
        \bigl( \tilde T (\la) \otimes \tilde T (\m) \bigr) =
        \bigl( \tilde T (\m) \otimes \tilde T (\la) \bigr)
        \check r (\la - \m) \qd.
\end{equation}
It follows that $\tr \bigl( \tilde T (\la) \bigr)$ can be diagonalized
by the algebraic Bethe ansatz if a pseudo vacuum exists on which
$\tilde T (\la)$ acts triangularly.

The operators $\tilde A (\la), \dots, \tilde D(\la)$ act on
$({\mathbb C}^2)^{\otimes N} \otimes \CH$. In this space let us define
the vector
\begin{equation}
     \hat \Om = \tst{{1 \choose 0}^{\otimes N}} \otimes \Om \qd.
\end{equation}
We shall see that $\hat \Om$ is an appropriate pseudo vacuum for
$\tilde T (\la)$. Let us denote the elements of the gl(2) standard
basis by $e_\a^\be$. Taking into account that
\begin{equation}
     r^{(0)}_{0, j} (\la) = \begin{pmatrix}
                               (b - a) \, {e_j}_1^1 - a \, {e_j}_2^2 &
			       b \, {e_j}_2^1 \\
			       b \, {e_j}_1^2 &
			       - a \, {e_j}^1_1 + (b - a) \, {e_j}^2_2 
                            \end{pmatrix} \qd,
\end{equation}
when written as a matrix in auxiliary space, and using the definition
(\ref{deftaux}) of the auxiliary monodromy matrix, we obtain
\begin{align}
     A^{(0)} (\la) \tst{{1 \choose 0}^{\otimes N}} & =
        \prod_{j=1}^N \bigl( b(\la - \la_j) - a(\la - \la_j) \bigr) \cdot
        \tst{{1 \choose 0}^{\otimes N}} \qd, \\
     D^{(0)} (\la) \tst{{1 \choose 0}^{\otimes N}} & =
        (-1)^N \prod_{j=1}^N a(\la - \la_j) \cdot
        \tst{{1 \choose 0}^{\otimes N}} \qd, \\
     C^{(0)} (\la) \tst{{1 \choose 0}^{\otimes N}} & = 0 \qd.
\end{align}
Hence, using (\ref{para}), (\ref{null}) and (\ref{abcdt}), we conclude
that
\begin{align} \label{aom}
     \tilde A (\la) \, \hat \Om & = (-1)^N a_2 (\la)
        \prod_{j=1}^N \frac{a(\la - \la_j)}{a(\la_j - \la)} \,
	\hat \Om \qd, \\ \label{dom}
     \tilde D (\la) \, \hat \Om & = (-1)^N a_3 (\la)
        \prod_{j=1}^N a(\la - \la_j) \, \hat \Om \qd, \\
     \tilde C (\la) \, \hat \Om & = 0 \qd.
\end{align}
This means that $\tilde T (\la)$ acts triangularly on $\hat \Om$.

Let us briefly recall the algebraic Bethe ansatz solution of the
gl(2) generalized model (see e.g.\ \cite{KBIBo}). Out of the 16 relations
contained in (\ref{gl2yba}) we pick out the following three,
\begin{align} \label{tbb}
     & \tilde B(\la) \tilde B(\m) = \tilde B(\m) \tilde B(\la) \qd, \\
       \label{tab}
     & \tilde A(\la) \tilde B(\m) = - \, \frac{b(\la - \m)}{a(\la - \m)}
          \, \tilde B(\la) \tilde A(\m) +
	  \frac{\tilde B(\m) \tilde A(\la)}{a(\la - \m)} \qd, \\
       \label{tdb}
     & \tilde D(\la) \tilde B(\m) = - \, \frac{b(\m - \la)}{a(\m - \la)}
          \, \tilde B(\la) \tilde D(\m) +
	  \frac{\tilde B(\m) \tilde D(\la)}{a(\m - \la)} \qd.
\end{align}
Moving $\tilde A(\la)$ and $\tilde D(\la)$ through a product
$\tilde B(\m_1) \dots \tilde B(\m_M)$ by means of (\ref{tbb})-(\ref{tdb})
leads to
\begin{align}
     & \tilde A(\la) \Bigl[ \prod_{k=1}^M \tilde B(\m_k) \Bigr] =
        \Bigl[ \prod_{k=1}^M \tilde B(\m_k) \Bigr] \, \tilde A(\la) \,
	\prod_{k=1}^M \frac{1}{a(\la - \m_k)} \nn \\ & \mspace{126.0mu}
	- \sum_{k=1}^M \Bigl[ \tilde B(\la)
	\prod_{l = 1 \atop l \ne k}^M \tilde B(\m_l) \Bigr] \,
	\tilde A (\m_k) \, \frac{b(\la - \m_k)}{a(\la - \m_k)}
	\prod_{l = 1 \atop l \ne k}^M \frac{1}{a(\m_k - \m_l)} \qd,
	   \\[1ex]
     & \tilde D(\la) \Bigl[ \prod_{k=1}^M \tilde B(\m_k) \Bigr] =
        \Bigl[ \prod_{k=1}^M \tilde B(\m_k) \Bigr] \, \tilde D(\la) \,
	\prod_{k=1}^M \frac{1}{a(\m_k - \la)} \nn \\ & \mspace{126.0mu}
	- \sum_{k=1}^M \Bigl[ \tilde B(\la)
	\prod_{l = 1 \atop l \ne k}^M \tilde B(\m_l) \Bigr] \,
	\tilde D (\m_k) \, \frac{b(\m_k - \la)}{a(\m_k - \la)}
	\prod_{l = 1 \atop l \ne k}^M \frac{1}{a(\m_l - \m_k)} \qd.
\end{align}
We add the latter equations and use the fact that $b(\la)/a(\la)$ is
odd. Then
\begin{align} \label{ttb}
     & \bigl( \tilde A(\la) + \tilde D(\la) \bigr)
        \Bigl[ \prod_{k=1}^M \tilde B(\m_k) \Bigr] =
        \Bigl[ \prod_{k=1}^M \tilde B(\m_k) \Bigr]
	   \nn \\ & \mspace{184.0mu} \cdot
	   \Bigl\{ \tilde A(\la) \prod_{k=1}^M \frac{1}{a(\la - \m_k)}
	         + \tilde D(\la) \prod_{k=1}^M \frac{1}{a(\m_k - \la)}
	   \Bigr\} \nn \\[1ex] & \qd
	+ \sum_{k=1}^M \Bigl[ \tilde B(\la)
	     \prod_{l = 1 \atop l \ne k}^M \tilde B(\m_l) \Bigr] 
	  \frac{b(\m_k - \la)}{a(\m_k - \la)}
	  \prod_{l = 1 \atop l \ne k}^M \frac{1}{a(\m_l - \m_k)}
	  \nn \\ & \mspace{184.0mu} \cdot
	\Bigl\{ \tilde A (\m_k)
	   \prod_{l = 1 \atop l \ne k}^M
	   \frac{a(\m_l - \m_k)}{a(\m_k - \m_l)} - \tilde D (\m_k)
	\Bigr\} \qd.
\end{align}
Now $\hat \Om$ is a joint eigenvector of $\tilde A(\la)$ and $\tilde
D(\la)$ (see (\ref{aom}), (\ref{dom})). Thus, acting with both sides
of equation (\ref{ttb}) on $\hat \Om$ we see that $\tilde B(\m_1) \dots
\tilde B(\m_M) \, \hat \Om$ is an eigenvector of $\tr \bigl( \tilde T
(\la) \bigr) = \tilde A(\la) + \tilde D(\la)$,
\begin{align} \label{ttbmult}
     & \tr \bigl( \tilde T (\la) \bigr) \,
        \tilde B(\m_1) \dots \tilde B(\m_M) \, \hat \Om =
        \tilde \La (\la) \,
	\tilde B(\m_1) \dots \tilde B(\m_M) \, \hat \Om \qd, \\
\intertext{with eigenvalue} \label{ttev}
     & \tilde \La (\la) = (-1)^N \prod_{j=1}^N a(\la - \la_j) \nn \\
        & \mspace{54.0mu} \cdot
        \Bigl\{ a_2 (\la) \prod_{j=1}^N \frac{1}{a(\la_j - \la)}
	   \prod_{k=1}^M \frac{1}{a(\la - \m_k)} +
	   a_3 (\la) \prod_{k=1}^M \frac{1}{a(\m_k - \la)}
        \Bigr\} \qd,
\end{align}
if the Bethe ansatz equations
\begin{equation} \label{tbae}
    \frac{a_2 (\m_k)}{a_3 (\m_k)}
       \prod_{j=1}^N \frac{1}{a(\la_j - \m_k)} =
       \prod_{l = 1 \atop l \ne k}^M
       \frac{a(\m_k - \m_l)}{a(\m_l - \m_k)}
\end{equation}
are satisfied for $k = 1, \dots, M$.

Since the vacuum subspace $\CH_0 \subset \CH$ is invariant under
$D(\la)$, it follows that $\tilde B(\m_1) \dots \tilde B(\m_M) \,
\hat \Om \in ({\mathbb C}^2)^{\otimes N} \otimes \CH_0$. In other words,
$\tilde B(\m_1) \dots \tilde B(\m_M) \, \hat \Om$ is a column vector
with $2^N$ rows having vectors in $\CH_0 \subset \CH$ as entries. Hence,
by the definition (\ref{h01}), (\ref{h02}) of the vacuum subspace,
$\tilde B(\m_1) \dots \tilde B(\m_M) \, \hat \Om$ is an eigenvector of
$A(\la)$, viewed as an operator on $({\mathbb C}^2)^{\otimes N} \otimes
\CH$,
\begin{equation} \label{atbmult}
     A(\la) \, \tilde B(\m_1) \dots \tilde B(\m_M) \, \hat \Om =
        a_1 (\la) \, \tilde B(\m_1) \dots \tilde B(\m_M) \, \hat \Om \qd.
\end{equation}
Recall that $B(\la_1) \otimes \dots \otimes B(\la_N)$ is a row vector
with $2^N$ columns. It follows from (\ref{tbmult}), (\ref{ttbmult})
and (\ref{atbmult}) that
\begin{multline} \label{baeve}
     \Bigl[ \bigotimes_{j=1}^N B(\la_j) \Bigr] \,
        \tilde B(\m_1) \dots \tilde B(\m_M) \, \hat \Om \\ =
	B_{i_1} (\la_1) \dots B_{i_N} (\la_N)
	\bigl[ \tilde B(\m_1) \dots \tilde B(\m_M)
	\hat \Om \bigr]^{i_1, \dots, i_N} \in \CH
\end{multline}
is an eigenvector of $t(\la)$ with eigenvalue
\begin{multline} \label{baeva}
     \La (\la) = a_1 (\la) \prod_{j=1}^N \frac{1}{a(\la_j - \la)} \\
                 - a_2 (\la) \prod_{j=1}^N \frac{1}{a(\la_j - \la)}
		             \prod_{k=1}^M \frac{1}{a(\la - \m_k)}
                 - a_3 (\la) \prod_{k=1}^M \frac{1}{a(\m_k - \la)} \qd,
\end{multline}
if the Bethe ansatz equations
\begin{equation} \label{bae}
     \frac{a_1 (\la_j)}{a_2 (\la_j)} = \prod_{k=1}^M
                                       \frac{1}{a(\la_j - \m_k)}
\end{equation}
are satisfied for $j = 1, \dots, N$. Thus we have obtained the
eigenvectors (\ref{baeve}) and eigenvalues (\ref{baeva}) of the transfer
matrix $t(\la)$ of the gl(1$|$2) generalized model. The eigenvalues and
eigenvectors depend on two sets of Bethe ansatz roots $\{\la_j\}_{j=1}^N$,
$\{\m_k\}_{k=1}^M$ determined by the Bethe ansatz equations (\ref{tbae})
and (\ref{bae}). We would like to stress that we did {\em not} specify
the functions $a_j (\la)$, $j = 1, 2, 3$, yet, nor did we require that
$D^1_2 (\la) \Om = 0$. Our calculation solely relied on the graded
Yang-Baxter algebra (\ref{gyba}) and on equations (\ref{para}) and
(\ref{null}).

\begin{remark}
Suppose all $\la_j$ and $\m_k$ be mutually distinct, and the parameters
$a_j (\la)$ have no singularities at $\la_j$, $\m_k$. Then $\La (\la)$,
equation (\ref{baeva}), has simple poles at $\la = \la_j$, $j = 1, \dots,
N$ and $\la = \m_k$, $k = 1, \dots, M$. The condition for the poles at
$\la = \la_j$ to vanish is
\begin{equation}
     \res \bigl\{ \La (\la_j) \bigr\} = 0 \qd, \qd j = 1, \dots, N
\end{equation}
and is equivalent to (\ref{bae}). Similarly, the equations
\begin{equation}
     \res \bigl\{ \La (\m_k) \bigr\} = 0 \qd, \qd k = 1, \dots, M
\end{equation}
are equivalent to (\ref{tbae}).
\end{remark}

\section{The fundamental representation}
In order to illustrate our results with an example let us reconsider
the algebraic Bethe ansatz of the supersymmetric $t$-$J$ model
\cite{EsKo92}. We have to connect the model to a representation of the
graded Yang-Baxter algebra (\ref{gyba}) and have to identify the
parameters $a_1 (\la)$, $a_2 (\la)$ and $a_3 (\la)$ of the
representation. The Bethe ansatz equations are then given by
(\ref{tbae}), (\ref{bae}), the eigenvectors and the eigenvalues of the
transfer matrix by (\ref{baeve}) and~(\ref{baeva}). We shall basically
follow the account of \cite{GoMu98,GoKo00}.

First of all, we introduce canonically anticommuting creation and
annihilation operators $c_{j, a}^+$, $c_{k, b}$ of electrons ($a, b =
\auf, \ab$; $j, k = 1, \dots, L$) and the Fock vacuum $|0\>$,
annihilated by all annihilation operators, $c_{k, b} |0\> = 0$. The
elements $(X_j)^\a_\be$, $\a, \be = 1, 2, 3$, of the matrix
\begin{equation} \label{xjtj}
     X_j = \begin{pmatrix}
	     (1 - n_{j \ab})(1 - n_{j \auf}) &
	     (1 - n_{j \ab}) c_{j \auf} &
	     c_{j \ab} (1 - n_{j \auf}) \\[.5ex]
	     (1 - n_{j \ab}) c_{j \auf}^+ &
	     (1 - n_{j \ab}) n_{j \auf} &
	     - c_{j \ab} c_{j \auf}^+ \\[.5ex]
	     c_{j \ab}^+ (1 - n_{j \auf}) &
	     c_{j \ab}^+ c_{j \auf} &
	     n_{j \ab} (1 - n_{j \auf})
           \end{pmatrix}
\end{equation}
form a complete set of projection operators on the space of states
locally spanned by the basis vectors $|0\>$, $c_{j \auf}^+ |0\>$,
$c_{j \ab}^+ |0\>$. Double occupancy of lattice sites is forbidden on
this space. Let ${X_j}_\a^\be = (X_j)^\a_\be$, $\a, \be = 1, 2, 3$. The
operator ${X_j}_\a^\a = 1 - n_{j \auf} n_{j \ab}$ projects the local
space of lattice electrons onto the space from which double occupancy is
excluded. The corresponding global projection operator is
\begin{equation} \label{notwo}
     P_0 = \prod_{j=1}^L (1 - n_{j \auf} n_{j \ab}) \qd.
\end{equation}

Owing to the fact that the operators ${X_j}_\a^\be$ are local projection
operators, it follows from general considerations \cite{GoMu98} that
the `$L$-matrix'
\begin{equation} \label{ltj}
     \CL_j (\la) = a(\la) I_3 + b(\la) \begin{pmatrix}
		    {X_j}_1^1 & {X_j}_2^1 & {X_j}_3^1 \\[.5ex]
		    {X_j}_1^2 & - {X_j}_2^2 & - {X_j}_3^2 \\[.5ex]
		    {X_j}_1^3 & - {X_j}_2^3 & - {X_j}_3^3
                 \end{pmatrix} \qd.
\end{equation}
is a representation of the graded Yang-Baxter algebra (\ref{gyba}).
This representation has been termed fundamental graded representation
in \cite{GoMu98}. The action of $\CL_j (\la)$ on the Fock vacuum 
obviously is
\begin{equation} \label{ltjvac}
     \CL_j (\la) |0\> = \begin{pmatrix}
		    1 & b(\la) {X_j}_2^1 & b(\la) {X_j}_3^1 \\[.5ex]
		    0 & a(\la) & 0 \\[.5ex]
		    0 & 0 & a(\la)
                 \end{pmatrix} |0\> \qd.
\end{equation}
The matrix $\CL_j (\la)$ generates the supersymmetric $t$-$J$ model at
a single site. The corresponding monodromy matrix of the $L$-site
model is
\begin{equation} \label{ttj}
     \CT (\la) = \CL_L (\la) \dots \CL_1 (\la) \qd.
\end{equation}
Its action on the Fock vacuum follows from (\ref{ltjvac}) as
\begin{equation} \label{ttjvac}
     \CT (\la) |0\> = \begin{pmatrix}
		    1 & B_1 (\la) & B_2 (\la) \\[.5ex]
		    0 & a^L (\la) & 0 \\[.5ex]
		    0 & 0 & a^L (\la)
                 \end{pmatrix} |0\> \qd.
\end{equation}
From the latter equation we can read off the parameters of the
representation, $a_1 (\la) = 1$, $a_2 (\la) = a_3 (\la) = a^L (\la)$.
And we are done (compare \cite{EsKo92}, equations (3.47), (3.48) and
(3.50)).

\begin{remark}
Note that the Hamiltonian of the supersymmetric $t$-$J$ model is
\begin{equation}
     H = - \i c \, \6_\la \ln \bigl\{ \bigl( \str (\CT (0)) \bigr)^{-1}
                                       \str (\CT (\la)) \bigr\}
				       \Bigr|_{\la = 0} \qd.
\end{equation}
Because it acts on the restricted space of electronic states, where no
lattice site is doubly occupied, we may replace it with (see
\cite{GoMu98,GoKo00})
\begin{equation}
     H P_0 = P_0 \Bigl\{ - \sum_{j=1}^L
        (c_{j, a}^+ c_{j+1, a} + c_{j+1, a}^+ c_{j, a})
	+ 2 \sum_{j=1}^L \left( S_j^\a S_{j+1}^\a
	- \frac{n_j n_{j+1}}{4} + n_j \right) \Bigr\} P_0
	\qd.
\end{equation}
\end{remark}

\section{Connection to Lieb-Wu equations}
We demonstrate that for an appropriate choice of the parameters of the
gl(1$|$2) generalized model the Bethe ansatz equations (\ref{tbae}),
(\ref{bae}) turn into the Lieb-Wu equations \cite{LiWu68}, which are the
Bethe ansatz equations of the Hubbard model.

Let us introduce two functions of the spectral parameter, $k(\la)$ and
$v(\la)$, defined by
\begin{equation} \label{lwspec}
     \sin k = \la \qd, \qd v = \la - \i c/2 \qd.
\end{equation}
Setting
\begin{align}
     & k_j = k(\la_j) \qd, \qd j = 1, \dots, N \qd, \\
     & v_k = v (\m_k) \qd, \qd k = 1, \dots, M
\end{align}
and using the explicit form (\ref{defab}) of the function $a(\la)$, we
can rewrite the Bethe ansatz equations (\ref{bae}), (\ref{tbae}) as
\begin{align}
     & \frac{a_1 (\sin k_j)}{a_2 (\sin k_j)} =
       \prod_{k=1}^M \frac{v_k - \sin k_j - \i c/2}
                          {v_k - \sin k_j + \i c/2} \qd, \\
     & \frac{a_2 (v_k + \i c/2)}{a_3 (v_k + \i c/2)}
       \prod_{j=1}^N \frac{v_k - \sin k_j - \i c/2}
                          {v_k - \sin k_j + \i c/2} =
       \prod_{l = 1 \atop l \ne k}^M \frac{v_k - v_l - \i c/2}
                                          {v_k - v_l + \i c/2} \qd,
\end{align}
for $j = 1, \dots, N$; $k = 1, \dots, M$. Obviously, the latter two
equations turn into the Lieb-Wu equations through the choice
\begin{equation} \label{lwpara}
     \frac{a_1 (\sin k)}{a_2 (\sin k)} = e^{\i k L} \qd, \qd
     \frac{a_2 (v + \i c/2)}{a_3 (v + \i c/2)} = 1
\end{equation}
of the parameters of the generalized model.

Similarly, replacing $\sin k$ with $k$ in equations (\ref{lwspec})-%
(\ref{lwpara}) we obtain the Bethe ansatz equations \cite{Yang67} of
Yang's model of electrons interacting via a delta function potential.

\section{Conclusions}
We have obtained the algebraic Bethe ansatz solution of the gl(1$|$2)
generalized model. The Bethe ansatz equations (\ref{tbae}), (\ref{bae})
and the eigenvalue of the transfer matrix (\ref{baeva}) depend on three
functional parameters, which are determined by the respective
representation of the graded Yang-Baxter algebra (\ref{gyba}). Choosing
the parameters appropriately we obtain the Bethe ansatz equations for the
models discussed in the introduction. In order to calculate the transfer
matrix eigenvalues for the models based on higher representations of
gl(1$|$2) we have to supply additional arguments based on the ideas of
fusion and analyticity of the eigenvalue (see \cite{PfFr96}).

What are the further applications of our result? They relate to points
(ii) and (iii) of the list in the introduction. We do not have much
doubt that Reshetikhin's work \cite{Reshetikhin86} on the norm of the
gl(3) generalized model carries over almost literally to the gl(1$|$2)
case (see point (ii) in the introduction). It is, however, point (iii)
which enthralls us most. We propose to systematically seek for a
representation of the graded Yang-Baxter algebra (\ref{gyba}) with
parameters $a_1 (\la)$, $a_2 (\la)$ and $a_3 (\la)$ satisfying
(\ref{lwpara}). Such a representation would certainly much improve our
understanding of the algebraic structure of the Hubbard model. It would
probably enable to prove the norm conjecture proposed in \cite{GoKo99b}
for the Bethe ansatz wave functions of the Hubbard model, and, combined
with the results of \cite{GoKo00}, might eventually lead to form factor
formulae for local operators in the finite Hubbard chain.

Unfortunately, for the case of gl(1$|$2) no theorem is known so far
that would guarantee the existence of a representation of the graded
Yang-Baxter algebra (\ref{gyba}) with arbitrary functional parameters.
It would be highly desirable to find out whether or not a generalization
of the work of Korepin and Tarasov \cite{Korepin82b,Tarasov84,Tarasov85}
to the gl(1$|$2) case is possible. As long as this question is not
settled it remains unclear whether the appearance of the Lieb-Wu
equations in the context of the gl(1$|$2) generalized model reflects
a deep connection with the Hubbard model or is just a coincidence.

\subsection*{Acknowledgments}
The author likes to thank V. E. Korepin and M. Shiroishi for helpful
discussions. He thanks K. Sakai for encouraging him to publish his
results.

\renewcommand{\thesection}{\Alph{section}}

\renewcommand{\theequation}{\thesection.\arabic{equation}}

\clearpage

\setcounter{section}{0}

\append{Graded algebras}
In this appendix we shall recall the basic concepts of graded vector
spaces and graded associative algebras. In the context of the quantum
inverse scattering method these concepts were first utilized by Kulish and
Sklyanin \cite{KuSk80,Kulish85}.

Graded vector spaces are vector spaces equipped with a notion of odd
and even, that allows us to treat fermions within the formalism of the
quantum inverse scattering method (see \cite{GoMu98,GoKo00}). Let us
consider a finite dimensional vector space $V$, which is the direct
sum of two subspaces, $V = V_0 \oplus V_1$, $\dim V_0 = m$,
$\dim V_1 = n$. We shall call $v_0 \in V_0$ even and $v_1 \in V_1$
odd. The subspaces $V_0$ and $V_1$ are called the homogeneous components
of $V$. The parity $\p$ is a function $V_i \rightarrow \mathbb{Z}_2$
defined on the homogeneous components of~$V$,
\begin{equation}
     \p(v_i) = i \qd, \qd i = 0, 1 \qd, \qd v_i \in V_i \qd.
\end{equation}
The vector space $V$ endowed with this structure is called a graded
vector space or super space. 

Let ${\cal A}$ be an associative algebra (with unity), which is graded
as a vector space. Suppose $X, Y \in {\cal A}$ are homogeneous. If the
product $XY$ is homogeneous with parity
\begin{equation} \label{homab}
     \p(XY) = \p(X) + \p(Y) \qd,
\end{equation}
then ${\cal A}$ is called a graded associative algebra \cite{KuSk80}.

For any two homogeneous elements $X, Y \in {\cal A}$ let us define the
super-bracket
\begin{equation} \label{superbracket}
     [X,Y]_\pm = XY - (-1)^{\p(X)\p(Y)} YX \qd,
\end{equation}
and let us extend this definition linearly in both of its arguments
to all elements of~${\cal A}$.

Let $p: \{1, \dots, n \} \rightarrow {\mathbb Z}_2$. The set of all
$n \times n$ matrices $A, B, C, \dots$ with entries in ${\cal A}$, such
that $\p (A^\a_\be) = \p (B^\a_\be) = \p (C^\a_\be) = \dots = p(\a) +
p(\be)$ is an associative algebra, say ${\rm Mat} ({\cal A}, n)$, since
$\p (A^\a_\be B^\be_\g) = p(\a) + p(\g)$. For $A, B \in {\rm Mat}
({\cal A}, n)$ we define the super tensor product (or graded tensor
product)
\begin{equation}
     (A \otimes_s B)^{\a \g}_{\be \de} = (-1)^{(p(\a) + p(\be)) p(\g)}
                                            A^\a_\be B^\g_\de \qd.
\end{equation}
This definition has an interesting consequence. Let $A, B, C, D \in
{\rm Mat} ({\cal A}, n)$, such that $[B^\a_\be, C^\g_\de]_\pm = 0$.
Then
\begin{equation} \label{abcd}
     (A \otimes_s B)(C \otimes_s D) = AC \otimes_s BD \qd.
\end{equation}

\append{Three identities}
In this appendix we provide three identities that are useful for the
proof of equation (\ref{dbmult}). The first one is
\begin{align} \label{id1}
     & \biggl\{ \Bigl[ \bigotimes_{j=1}^N B(\la_j) \Bigr]
        \otimes D(\la) \biggr\} \, \check r^{(0)}_{N-1, N} (\la -\la_N)
	\check r^{(0)}_{N-2, N-1} (\la - \la_{N-1}) \dots
	\check r^{(0)}_{0, 1} (\la - \la_1) \nn \\
     & \qd = \biggl\{ \Bigl[ \bigotimes_{j=1}^N B(\la_j) \Bigr]
        \otimes I_2 \biggr\}
	\Bigl\{ I_2^{\otimes N} \otimes D(\la) \Bigr\} \,
	P_{N-1 N} \dots P_{12} P_{01} \nn \\[-2ex]
	& \mspace{324.0mu}
	\cdot r^{(0)}_{0, N} (\la -\la_N) \dots
	r^{(0)}_{0, 1} (\la - \la_1) \nn \\
     & \qd = \biggl\{ I_2 \otimes
        \Bigl[ \bigotimes_{j=1}^N B(\la_j) \Bigr] \biggr\}
	\, D^{(0)}_0 (\la) T^{(0)} (\la) \qd.
\end{align}
The second one is
\begin{align} \label{id2}
     & \biggl\{ B(\la) \otimes
        \Bigl[ \bigotimes_{k=1 \atop k \ne j}^N B(\la_k) \Bigr]
        \otimes D(\la_j) \biggr\}
	\, \check r^{(0)}_{N-1, N} (\la_j -\la_N) \dots
	\check r^{(0)}_{j, j+1} (\la_j - \la_{j+1}) \nn \\
     & \qd = \biggl\{ I_2 \otimes B(\la) \otimes
        \Bigl[ \bigotimes_{k=1 \atop k \ne j}^N B(\la_k) \Bigr]
	\biggr\} \, P_{01} \dots P_{N-1 N}
	\Bigl\{ I_2^{\otimes N} \otimes D(\la_j) \Bigr\} \nn \\[-3ex]
	& \mspace{252.0mu}
	\cdot \check r^{(0)}_{N-1, N} (\la_j -\la_N) \dots
	\check r^{(0)}_{j, j+1} (\la_j - \la_{j+1}) \nn \\[1ex]
     & \qd = \biggl\{ I_2 \otimes B(\la) \otimes
        \Bigl[ \bigotimes_{k=1 \atop k \ne j}^N B(\la_k) \Bigr]
	\biggr\} \, P_{01} \dots P_{j-1 j} \nn \\[-3ex]
	& \mspace{252.0mu}
	\cdot D_j^{(0)} (\la_j) \,
	r^{(0)}_{j, N} (\la_j - \la_N) \dots
	r^{(0)}_{j, j+1} (\la_j - \la_{j+1}) \nn \\[1ex]
     & \qd = \biggl\{ I_2 \otimes B(\la) \otimes
        \Bigl[ \bigotimes_{k=1 \atop k \ne j}^N B(\la_k) \Bigr]
	\biggr\} \, P_{01} S_{j-1}^{(0)} \,
	r^{(0)}_{j, j-1} (\la_j - \la_{j-1}) \cdot \nn \\[-1ex]
	& \mspace{144.0mu}
	\dots r^{(0)}_{j, 1} (\la_j - \la_1)
	D_j^{(0)} (\la_j)
	r^{(0)}_{j, N} (\la_j - \la_N) \dots
	r^{(0)}_{j, j+1} (\la_j - \la_{j+1}) \nn \\[1.5ex]
     & \qd = \biggl\{ I_2 \otimes B(\la) \otimes
        \Bigl[ \bigotimes_{k=1 \atop k \ne j}^N B(\la_k) \Bigr]
	\biggr\} \, P_{01} S_{j-1}^{(0)}
	\Bigl\{ I_2 \otimes
	\tr_0 \bigl( D_0^{(0)} (\la_j) T^{(0)} (\la_j) \bigr)
	\Bigr\}\ .
\end{align}
The third identity is an identity between rational functions that can
be easily proven by means of Liouville's theorem,
\begin{multline} \label{id3}
     \frac{b(\la - \la_{N+1})}{a(\la - \la_{N+1})}
     \prod_{k=1}^N \frac{1}{a(\la_{N+1} - \la_k)} \\
        = \frac{b(\la - \la_{N+1})}{a(\la - \la_{N+1})}
	  \prod_{j=1}^N \frac{1}{a(\la - \la_j)}
        - \sum_{j=1}^N \frac{b(\la_j - \la_{N+1})}{a(\la_j - \la_{N+1})}
                       \frac{b(\la - \la_j)}{a(\la - \la_j)}
	  \prod_{k=1 \atop k \ne j}^N \frac{1}{a(\la_j - \la_k)} \qd.
\end{multline}
The induction step in the proof of (\ref{dbmult}) may be done as follows.
First, insert (\ref{id1}) and (\ref{id2}) into (\ref{dbmult}). Second,
take the tensor product of the resulting equation with $B(\la_{N+1})$.
Third, use (\ref{zamo}) and (\ref{db}) to bring the operators into the
required order. Finally, use (\ref{id3}) to rearrange the resulting
terms in the desired way.

%\bibliographystyle{phys}
%\bibliography{hub}

\end{document}